\documentclass[aps,prl,twocolumn,groupedaddress,amsmath,amssymb]{revtex4}

\usepackage{graphicx}
\usepackage{dcolumn}
\usepackage{bm}

\begin{document}

\title{Analytic Theory of Wannier-Stark Quantization in Two Dimensions}
\author{Alexander Onipko}
\affiliation{Laboratory on Quantum Theory in Link{\"o}ping, 
http://www.lqtl.org, 587 35 Link{\"o}ping, Sweden}
\email[]{onipko.alexander@bredband.net}
\author{Lyuba Malysheva}
\affiliation{Bogolyubov Institute for Theoretical Physics, 03680 Kyiv,
Ukraine}

\date{\today} 

\begin{abstract}
For the first time, one-particle theory of Wannier-Stark quantization for a $a(N$$-$1)-long chain affected by a homogeneous electric field $E_e$ is extended to the 2D case of 
$L_N$=$a(N$$-$1)-long and $L_{\cal N}$=$a$$({\cal N}$$-$1)-wide 
conductor, which is modeled by the atomic square lattice with the electron site-energy change from atom to atom in the direction parallel to $\cal N$ axis by the amount of electric field parameter (efp) $\varepsilon \equiv aeE_e/|t| $ in units of $|t|$. It is shown that each field-provoked $\mu$-level in the chain spectrum gives birth to the $\mu$-subband of field-independent levels due to the electron-transfer interaction in the direction perpendicular to $E_e$. The level spacing and hence, the width of $\mu$-subbands $E^0_{\rm bw}$, is dictated by the conductor length, the hopping integral $t$, and by the lattice constant $a$.  Another principal result is that the levels, which are within the energy interval $0.5E^0_{\rm bw}$ on either sides of  the spectrum (that is the edge spectrum), correspond to the delocalized states which are extended in the direction perpendicular to the electric field. In this direction, the electron spectrum width $E^{\perp}_{\rm bw}$ is larger by $E^0_{\rm bw}$ than the spectrum width in the field direction $E^{\parallel}_{\rm bw}$. Several special cases of the interrelation between the electron energy versus applied voltage are identified, when the spacing between the $\mu$-subband centers is equal either to integer or fraction ($>1$) portion of dimensionless 
$\varepsilon$. Otherwise, the level spacing is shown to be irregular and, depending on 
$N$, $\cal N$,  and $\varepsilon$, it can be equal to any portion of $\varepsilon$. It is argued that one of straightforward applications of the presented theory is the quantum-mechanical explanation of Hall effect in two dimensions. 
\end{abstract}

\maketitle

\section{Introduction}

In 1960, Wannier introduced the concept of electron energy quantization in solids subjected to a constant homogeneous electric field \cite{Wannier,Wannier1}. In its essence, his concept was formulated for an infinite monoatomic chain described in the Wannier tight-binding approximation. It can be considered as the theory of Stark effect for a chain of interacting single-level atoms. Therefore, it was christened by the name Wannier-Stark ladder or WS quantization of electron energy, $E_\mu=\mu\varepsilon$, $\mu$ is an integer. The field parameter $\varepsilon$ ($-\varepsilon$) determines the change of the electron potential energy from one atom to the next along (against) the field.

Significant clarifications of the Wannier concept were suggested in works
\cite{H-O,Sai,S-G,Fuku,YaG} and some others, see quoted papers for additional references. In particular, it was shown that the edge of the spectrum of the field-affected {\it semi-infinite} tight-binding chain exhibits remarkably different regularity in the interlevel spacing. The polynomial representation of the exact solution of the spectral problem for the field-affected, $\cal N$-{\it atom long} tight-binding chain \cite{Lyuba} opened a window for deeper insight into the nature of Wannier-Stark quantization (WSQ). A number of accurate explicit expressions was derived which elucidated a rich variety of electric field effects on the chain electron spectrum \cite{SSC,prb63,NYAS}. The spectrum's gross structure that comes to light from the mathematically rigorous description of the finite monoatomic chain subjected to the electric field is briefly outlined next.

The difference in the electron potential energy $eV=eE_eL_{\cal N}$ between chain ends
widens the electron spectrum and splits it into three subbands: the midband,
$-|E^0_{\rm bw}-eV|/2\le E\le |E^0_{\rm bw}-eV|/2\equiv E^{\rm b}_{\rm els}$, and two tilted bands below and above the midband, Fig. 1a. If $eV<E^0_{\rm bw}$, all levels (if any) within tilted bands correspond to end-localized states (els), the eigen energies within the midband correspond to the extended states (es), upper diagram. The increase of $eV$ to the value of $E^0_{\rm bw}$ narrows the es-band width to zero. The tilted bands come in top-to-bottom touch one with another, and the tilted bands reach their maximal width equal to the zero-field band width  
$E^0_{\rm bw}$, mid diagram. Further increase of $eV$ opens the band labeled as WSs (Wannier-Stark states, lower diagram). All levels within this band correspond to the electron states which are predominantly localized either near the left chain end or near the right chain end.

In the classic description of free particle motion, the potential energy profile 
has the parallelogram shape (shown in Fig. 1a by thick solid vertical lines and thin solid sloped lines). Within the area of upper (lower) els-band, the particle moves in the right triangle (in the base-up right triangle) hard-wall potential; within the area of es-band, the particle moves in the rectangular hard-wall potential; and within the area of WSs-band, the particle experience the parallelogram-shaped hard-wall potential. With the increase of $eV$, the sloped  sides of the parallelogram and triangles become longer, whereas the horizontal lines indicating the borders between WSs- and els-bands, become shorter. In the limit $eV\rightarrow\infty$  (${\cal N}$ = const), the potential profile degenerates into an infinite vertical straight line with no place for the classic particle. This observation gives emphasis to the purely quantum nature of the WS effect in solids. In the limit ${\cal N}\rightarrow\infty$ ($eV$ = const), $E_{\rm bw}=E^0_{\rm bw}$.  

The refinements of the Wannier concept seem to be frozen in one-dimensional world of linear molecules and superlattices since year 2001. In this paper, we report an explicit expression of the field-affected electron spectrum derived from the exact characteristic equation of Hamiltonian matrix for $N$$\times$${\cal N}$ square atomic lattice. It gives a fascinating picture of WSQ having obvious parentage of WSQ in one dimension and, at the same time, exhibiting many prominent distinctions with new physics in the background.

\section{Analytic Solution of the Eigenvalue Problem}

In the tight-binding, nearest-neighbor approximation, 
the electron wave function for the atomic lattice shown in Fig.~\ref{Fig2} can be written in the form of expansion
\begin{align}\label{1}
\Psi_{\rm 2D} &= \sqrt{\dfrac{2}{N+1}}\sum_{n=1}^N 
\sum_{ {\bar m}=-({\cal N}-{\tilde N}) }^{{\tilde N}-1}
\sin(k_nn)\phi_{\bar m}|n,{\bar m}\rangle,\notag \\ {\tilde N} & \equiv 
[({\cal N}+1)/2],
\end{align}
where $k_n=\pi n/(N+1)$, $n$ = 1,2,\dots,$N$. Here and in what follows, the square brackets indicate the integer part of the argument. The one-dimensional counterpart of Eq. (1) is
$
\Psi_{\rm 1D} = \sum_{ {\bar m}=-({\cal N}-{\tilde N}) }^{{\tilde N}-1}
\psi_{\bar m}|n,{\bar m}\rangle. 
$

\begin{figure}[t!]
\includegraphics[width=0.45\textwidth,height=0.32\textheight]{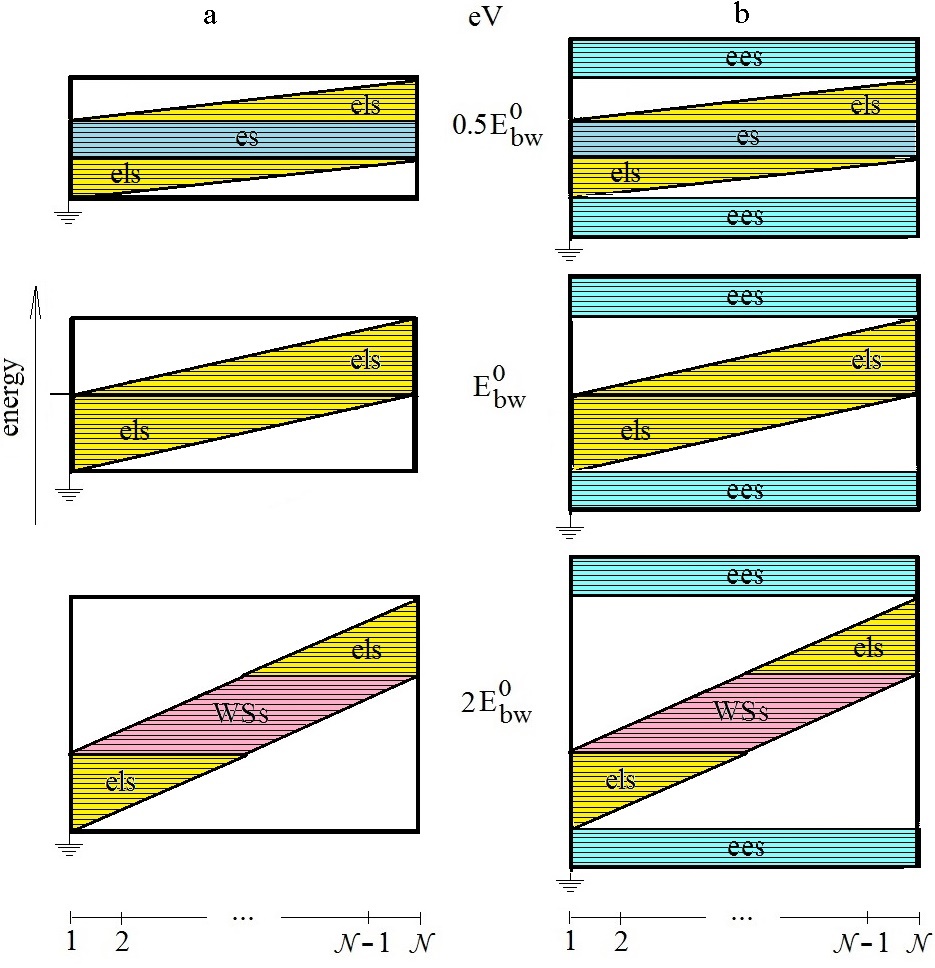}
\caption{
The spectrum's gross structure of $\cal N$-atom long chain (a) and $\cal N$$\times$$N$ 
($N$-long and $\cal N$-wide)  atomic square lattice (b) for particular values of the electric potential $V$. Thick solid frames indicate the hard wall potential experienced by a particle inside the walls. Thin-solid horizontal lines indicate the borders between the tilted bands of els-levels  corresponding to the  states, which are preferably localized near the lower or upper chain ends (a) [lattice edges (b)] and the midband of extended states (a,b) [edge extended states -- ees (b)], and Wannier-Stark states (WSs). The latter are localized between the sloped lines. The $\cal N$-$E$ areas of the existence of different types of states are distinguished by colors: yellow (els), dark  blue (es),  light blue (ees), and rose (WSs). White color indicate the areas which are inaccessible for classic particles. The width of electron spectrum $E^{\parallel}_{\rm bw}$ is equal to $E^0_{\rm bw}+eV$ (a); the width of electron spectrum $E^{\perp}_{\rm bw}$ is equal to $2E^0_{\rm bw}+eV$ (b).
}
\label{Fig1}
\end{figure}

The expansion coefficients $\phi_{\bar m} $  
must satisfy the Schr{\"o}dinger equation $H\Psi = E\Psi$, that is the system of equations
\begin{align}\label{2}
\left[E +\varepsilon \left ({\tilde N}-{\bar m}\right )+2\cos k_n\right]\phi_{\bar m}
& =-\left(\phi_{{\bar m}-1}+\phi_{{\bar m}+1}\right),  \notag \\
 \phi_{-({\cal N}-{\tilde N}+1)} &= \phi_{{\tilde N}} =0.
\end{align}

In the above equations, the electron site-energy at zero potential  (the Fermi energy) is set equal to zero, the energy is expressed in units of the hopping integral absolute value 
$|t|$, e.g., $E^0_{\rm bw}=4$, if ${\cal N}>>1$. For the sake of convenience, the choice of atom numbering along $\cal N$ axis preserves the electron spectrum symmetry with respect to the Fermi energy for any value of $eV$. 

The determinant of matrix ${\bar E}I -H$, $I_{n,{\bar m}}= \delta_{n,{\bar m}}$,
\begin{equation}\label{3}
H_{{\bar m},{\bar m}'}=\varepsilon({\tilde N}-{\bar m})\delta_{m,{\bar m}'}
-\delta_{|{\bar m}-{\bar m}'|,1}
\end{equation}
can be represented in the following form \cite{Lyuba}:
\begin{align}\label{4}
D_{\cal N}({\bar E}) &=J_{\nu +({\cal N} +1)/2}(z)Y_{\nu -({\cal N} +1)/2}(z)
\notag \\
&-Y_{\nu +({\cal N} +1)/2}(z)J_{\nu -({\cal N} +1)/2}(z),
\end{align}
where ${\bar E}\equiv E +2\cos k_n$, $\nu \equiv {\bar E}/{\varepsilon}$, $z \equiv 2/{\varepsilon}$, and $J_{\nu}(z)$ and $Y_{\nu}(z)$ are the Bessel functions of the first and second kind, respectively. 

The solutions to equation $D_{\cal N}(E)=0$ denoted by $E_\mu$, $\mu \! =\! 0,1,\dots,
{\cal N}\!-\!1$, give the electron eigen energies in ${\cal N}$-long atomic chain \cite{prb63}. From the form of $D_{\cal N}(\bar E)=0$, it is seen that each $\mu$-level in the chain spectrum splits into the $\mu$th subband of field-independent levels with $E_\mu$ in the subband center. This changes the level spacing possessed by the chain spectrum radically. As a consequence, the spectrum's gross structure in the 
$\cal N$-$E$ plane, including the spectrum width, also changes, compare Figs.\ref{Fig1}a and \ref{Fig1}b. Self-obvious, the overall spectrum structure in the $N$-$E$ plane (not shown) differs from that shown in Fig. \ref{Fig1}b, as does any other spectrum cross-section in the $N$$\cos\theta$-$E$ plane, $0<\theta<\pi/2$. 

In view of future applications of this model, we focus our attention on the details of the spectrum's gross structure shown in Fig.~\ref{Fig1}b. A few preliminary comments are in order. First, the dependence of the level spacing $\Delta$ on $N$ (in addition to its dependence on $V$ and $\cal N$) makes any $\Delta$ possible to appear in the electron spectrum. Therefore, we will concentrate the discussion on the spectrum regularities regarding the 
$\mu$-subband centers. Second, it is worth emphasizing that distinct from the levels of end-localized states, which form the edges of the atomic chain spectrum,  the edge spectrum of the atomic lattice is formed by the levels of edge extended states (ess). The difference of the nature of edge states in one and two dimensions, as well as the field-dependent difference between the $E$-$N$ and $E$-$\cal N$ cross-sections that associates with the non-equivalence of quasi-particle kinematics and dynamics along the field-parallel and field-perpendicular directions, gives rise to a number of spectrum peculiarities to be discussed.
 
\begin{figure}[t!]
\includegraphics[width=0.35\textwidth]{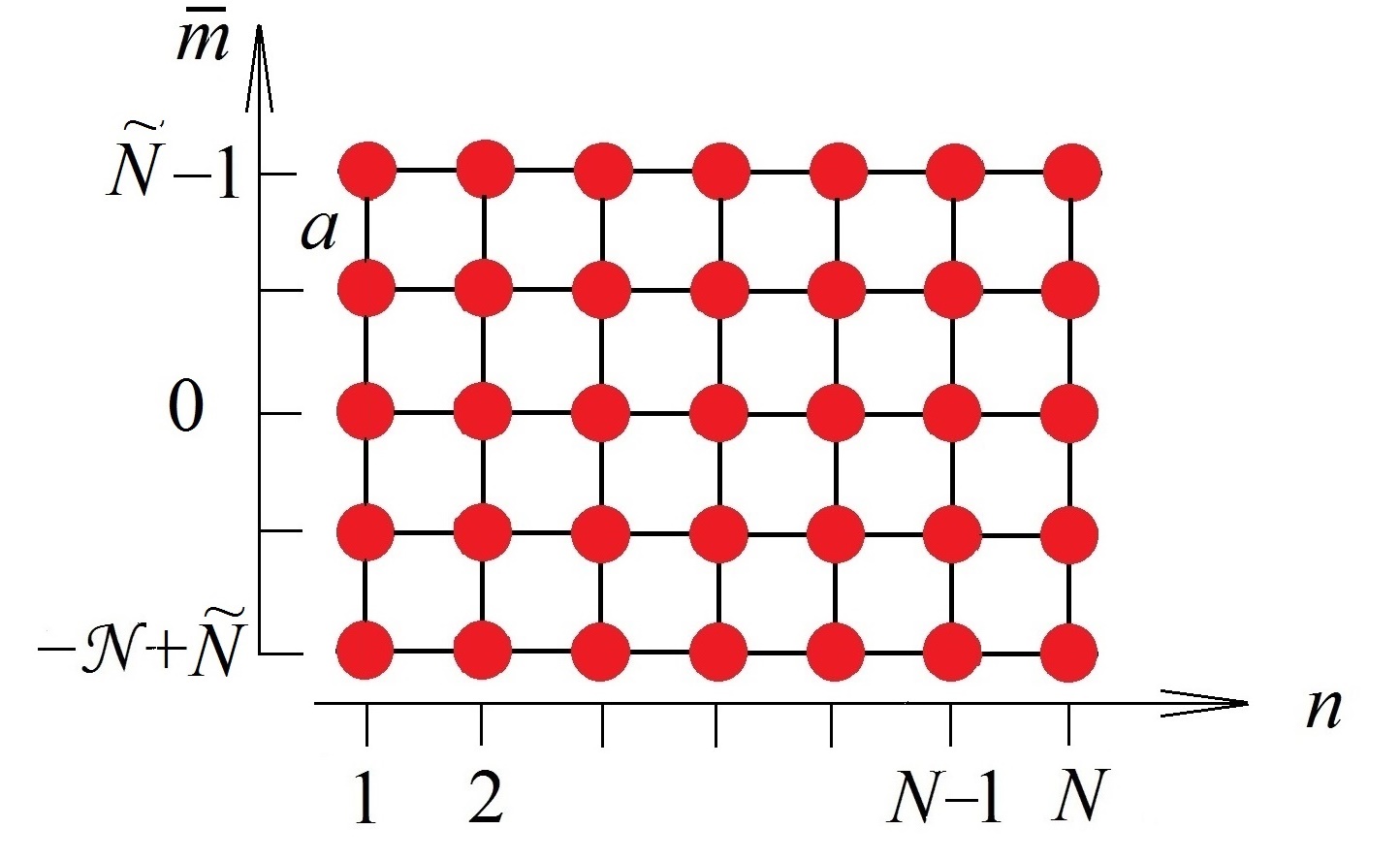}
\caption{
$N$$\times$$\cal N$ atomic lattice, where the electron site-energies are
evenly shifted in ${\bar m}$ direction by the value of  $\varepsilon$. The site
numbering from lower edge up is
${\bar m}$ = $-({\cal N}-{\tilde N})$, $\dots,-1,0,1,\dots$, ${\tilde N}-1$; 
$eV|_{{\tilde N}-1}-eV|_{-({\cal N}-{\tilde N})}=\varepsilon({\cal N}-1)=\!eV$.
}
\label{Fig2}
\end{figure}

In the ongoing analysis, the coexistence of edge-localized states, es- and WS-states at different voltages (as defined in Introduction) is discussed in a step-by-step manner. From now and on, $\cal N$ is supposed to be an odd number. The explicit expressions for the positive part of the spectrum are derived under the  restrictions ${\cal N} >>1$ and 
$\varepsilon << 1$ to make the obtained results relevant to the most cases of interest. Notice, the smallness of efp does not necessarily implies that $eV<<1$. 
We skip the analyses of small values of $N$, $\cal N$, or both, as the respective models do not seem to be relevant to any experiments in the visible future.

\begin{figure*}[t!]
\includegraphics[width=0.79\textwidth]{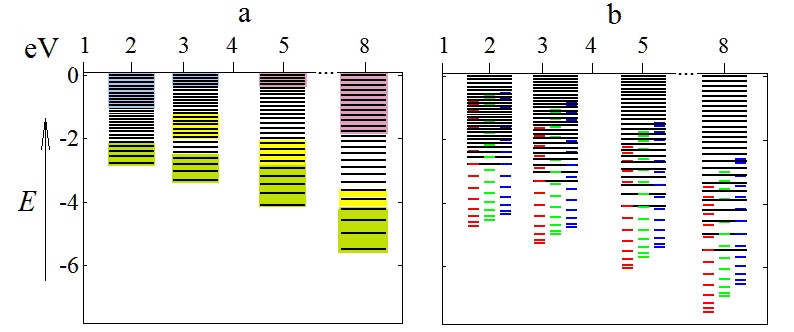}
\caption{Schematic illustration of the valence band spectrum (mirror reflection of the conduction band spectrum with respect to the zero-field Fermi energy) of $N$$\times$$\cal N$ atomic lattice. Horizontal lines, both longer and shorter, correspond either to 
$\mu$-levels calculated for $N$=1 (a), or to the $\mu$-subbands centers calculated for $N=15$, $\cal N$=51 (b). The spectra shown are calculated for (from left to the right) $eV$ = 2, $4\cos(\pi/5)$$\approx$3.24, 5, 8. The es-bands at different voltages are indicated by light blue color. For $eV=2$, 3$\varepsilon$ level spacing is valid excluding $\varepsilon$-small intervals near es-band edges. WS-bands are rose; for $eV$=5 ($\varepsilon$=0.1) and $eV$=8 ($\varepsilon$=0.16), the WS quantization
$\mu\varepsilon$ works well. The els-bands:  2$\varepsilon$-spaced levels (yellow part), irregularly-spaced levels (white part) separate the Airy spectrum (green) from the es- or WS-band. Yellow color: levels  close to ${\bar E}_{1,{\mu}}$=$eV/2$, see Eq. (\ref{15}). Cases  $eV=2$ and $eV$=$4\cos(\pi/5)$ are explained in the text. For  $N=15$ (b), only three lowest subbands are shown and shifted for clarity.
}
\label{Fig3}
\end{figure*}

\section{Field-Affected Electron Spectrum of 2D Atomic Lattice.}

{\bf Midband. Extended States. Case:\\ 
\underline{$eV< E^0_{\rm bw}$, 
$0\le{\bar E}<(E^0_{\rm bw}-eV)/2-\varepsilon$.}}
The excluded ${\varepsilon}$-interval contains at most one energy level. For the rest of half midband, the characteristic equation of the $\cal N$th order, equivalently, equation 
$D_{\cal N}({\bar E})=0$ can be represented in the form of a pair of interrelated transcendental equations
\begin{equation}\label{5}
{\cal D}^{\rm es}_{\cal N}(\xi^{\rm es})  =0, \quad
 {\bar E}=2 \cos \xi -\dfrac{eV}{2} = 2 \cos \xi'+\dfrac{eV}{2},
\end{equation}
where  
\begin{equation}\label{6}
 {\cal D}^{\rm es}_{\cal N}(\xi^{\rm es})\approx \frac{\varepsilon}{\pi}
\sqrt{\frac{1}{ \sin \xi \sin \xi'}}\sin[({\cal N}+1)\xi^{\rm es}],
\end{equation}
and
\begin{align}\label{7}
&({\cal N}+1)\xi^{\rm es} \notag \\ &= \xi+\dfrac{2}{\varepsilon}(\xi \cos \xi-\sin \xi)+
\xi'+\dfrac{2}{\varepsilon}(\sin \xi' -\xi' \cos \xi').
\end{align}
The roots of Eq. (\ref{5}) are $\xi^{\rm es}_\mu=\pi\mu/({\cal N}+1)$, where $\mu$ takes integer values 
${\tilde N}, {\tilde N}-1, \dots, {\tilde N}-N^{\rm es}+1$. The minimal value of $\xi^{\rm es}$ determines  the number of states $N^{\rm es}$ in a half of es-band of 
$\cal N$-long chain. We will return to this point later, see Eq. (\ref{13}) and on.

The classification of electron states by the values of renormalized quasi wave vector 
{\bf k} = $(k_n,\xi^{\rm es})$, as it is defined above, fits the central part of es-band. But this is not the end of the story.  It turns out that for the electron energies 
${\bar E}_{n,\mu}=\mu\varepsilon$, $\mu<<{\cal N}$, the determinant (\ref{4}) is accurately approximated by the expression
\begin{equation}\label{8}
{\cal D}_{\cal N}({\bar E}_{n,\mu})   \approx  (-1)^{{\tilde N}+\mu} 
\displaystyle\frac{\varepsilon}{\pi}\displaystyle
\frac{\sin\{2\mu \arccos(eV/4)\}}{\sin\{\arccos(eV/4)\}}.
\end{equation}
This equation tells us that the levels with $E_{n,\mu}=\mu\varepsilon-2\cos k_n$  can appear in the es-band occasionally, with some particular values of $\mu$. 
 
Further, for the values of $eV_\mu$ satisfying equation 
$
2\mu\arccos \left(eV_\mu/4 \right) = \pi \bar\mu, \quad
\bar\mu=0, 1, 2, \dots,
$
and, therefore, equation ${\cal D}_{\cal N}({\bar E}_{n,\mu}) =0$, we have
\begin{equation}\label{9}
eV_\mu =4\cos\left (\pi m'/m\right),
\end{equation}
where $m=3$, 4, $\dots$, and $m'=1$, 2, $\dots$, $<m/2$ are relatively prime numbers. For example, for $m=7$, $m'=1, 2, 3$. 

The pair of equations (\ref{8}), (\ref{9})  predicts the series of subbands
with the  subband centers at $E_\mu$, $\mu=0, m, 2m, 3m,\dots$, if $m$ is odd (odd series), and $\mu=0, m/2, m, 3m/2,\dots$ if $m$ is even (even series). 

Combining Eqs. (\ref{8}) and (\ref{9}), we found
\begin{equation}\label{10}
E_{n,\mu} = \frac{\mu\varepsilon}{1-2m'/m} -2\cos k_n, 
\quad\mu=0,1,2,\dots,
\end{equation}
where $m$ and $m'$ have the same meaning as above. This equation reveals the  interrelation between $E_{n,\mu}$ and the applied voltage, 
$V_\mu=\varepsilon({\cal N}-1)/e$ at fixed $\cal N$.  It refers to the middle part of the spectrum of $N$$\times$$\cal N$ atomic lattice. Large $\cal N$ are required for Eq. (\ref{10}) to be valid. In all other cases, it holds for arbitrary values of $\cal N$ and $N$. This suggests direct tests of  relevance of Eq. (\ref{10}) to the related experimental results on electric and magnetic effects in 2D materials.

Several examples of manifestation of such spectra are presented in Fig.~\ref{Fig3}. For  
$eV=0.5E^0_{\text bw}$ (left column), Eq. (\ref{9}) gives $m=3$ and $m'=1$. Then, from Eq. (\ref{10}),  $|E_{n,{\mu}+1}-E_{n,{\mu}}| \approx 3\varepsilon$. The exact calculations confirms that at this value of applied voltage, the es-band shows up the odd series as the triple-$\varepsilon$ spacing between the $\mu$-subbands. 

In the next example,  $eV$=$4\cos(\pi/5)$, $m=5$ and  $m'=1$. Thus,  the 
subband-center spacing is close to $(5/3) \varepsilon$.  In Fig.~\ref{Fig3}, the corresponding  part of the spectrum is blue.

{\bf Midband. Wannier-Stark States. Case:}\\
\underline{$eV> E^0_{\rm bw}$, $0<{\bar E}< E^{\rm b}_{\rm els}$}.
If $eV=E^0_{\rm bw}$, there is only one subband of extended states, 
$E_{n,0}= -2\cos k_n$. For larger values of $eV$, the mid-subband levels satisfying equation ${\bar E}_{n_{\rm c},\mu}=\mu\varepsilon$ ($k_{n_{\rm c}}=\pi/2$), must obey equation 
${\cal D}_{\cal N}(E_{n_{\rm c},\mu})=0$, where 
\begin{align}\label{11}
{\cal D}_{\cal N}&(E_{n_{\rm c},\mu})
\approx
(-1)^{{\tilde N}+\mu}\,\frac{{\varepsilon}}{\pi}
\frac{\sinh\{2\mu\cosh^{-1}(eV/4)\}}{\sinh\{\cosh^{-1}(eV/4)\}}.
\end{align}
According to \cite{prb63}, neglecting exponentially small  corrections we can write 
${\bar E}_{n_{\rm c},\mu}={\mu}{\varepsilon}$. This result specifies the range of validity and the meaning of WSQ in two dimensional atomic lattices: The spacing between the centers of field-independent bands is quantized as $\Delta_{\rm c}=\mu\varepsilon$. 

{\bf Tilted Band. Case:} 
\\
\underline{$E^{\rm b}_{\rm els} +\varepsilon<{\bar E}<\Big ( (E^0_{\rm bw}+eV)/2$} \underline{ $\equiv $  $ E^{\rm t}_{\rm els} \Big ) -\varepsilon$}. 
Within the indicated energy interval, determinant (\ref{4}) can be rewritten in the form
\begin{align}\label{12}
{\cal D}_{\cal N}({\bar E})\approx \frac{\varepsilon}{\pi}
\sqrt{\frac{1}{ \sin \xi' \sinh \delta}}
\sin\xi^{\rm els}\exp\left(\frac{2\Phi_\delta}{\varepsilon}+\delta\right),
\end{align}
where $\xi^{\rm els} \equiv 2\left( \sin\xi'-\xi' \cos\xi' \right)/\varepsilon+ \xi'+\pi/4$,
$2 \cosh \delta={\bar E} +eV/2$, $\Phi_\delta \equiv \delta\cosh \delta - \sinh \delta$. This expression of the determinant follows from Eq. (\ref{6}) with $\xi$ replaced by 
$i\delta$. 

Zeros of determinant (\ref{12}) are given by the roots of equation
$\sin\xi^{\rm els}=0$, or by
\begin{equation}\label{13}
\xi^{\rm els} =\pi\mu, \quad   \mu = 1,2,\dots ,N^{\rm els},
\end{equation}
where  $N^{\rm els}$=
$\left[\xi^{\rm els}|_{{\bar E}=0.5|E^0_{\rm bw}-eV|}/\pi\right]$. The number  
$ N^{\rm els}$  determines the distribution of states between es-, els-, and WSs-bands, since 
${\cal N} =$
$
2(N^{\rm es}+N^{\rm els})-\delta_{{\cal N}+1,2{\bar N}}$ and 
${\cal N} =
2(N^{\rm WSs}+N^{\rm els})-\delta_{{\cal N}+1,2{\bar N}}
$
that is also valid for even $\cal N$. 

At least in three cases, the solutions (\ref{13}) are qualitatively different.\\
\underline{{\bf Near the top of upper els-band}}. For
$E^{\rm t}_{{\rm els}}-{\bar E}<<1$, we get
\begin{equation}\label{14}
E_{n,\mu}=E^{\rm t}_{\rm els}  +\varepsilon -
\left \{ \dfrac{3}{2}\pi\left (\mu-\dfrac{1}{4} \right ) \varepsilon\right \}^{2/3}
-2\cos k_n,
\end{equation}
where the largest value of $\mu_{\rm max}$ is restricted by the condition
$\varepsilon<<1$. It is seen that the edge spectrum 
($E^{\rm t}_{\rm els}\le \bar E\le E^{\rm t}_{\rm els}+0.5E^0_{\rm bw}$) is formed by a set of mapped on each other cosine subbands. The highest (lowest) level in the edge spectrum is equal to $+(-)E^{\rm t}_{\rm els}+(-)0.5E^0_{\rm bw}$.

\underline{{\bf  For the energies satisfying}}
$|{\bar E}-eV/2|<<1$, and $eV>E^0_{\rm bw}$, we get
\begin{equation}\label{15}
{\bar E}_{n,\mu}=\dfrac{eV}{2}+2\varepsilon\left ( \dfrac{2}{\pi\varepsilon}+\dfrac{3}{4}-\left [\dfrac{2}{\pi\varepsilon}+\dfrac{3}{4} \right ] \right)+2\mu\varepsilon,
\end{equation}
where $\mu=0,\pm 1, \pm 2,\dots$, $|\mu|<<(2\varepsilon)^{-1}$.
Thus, this part of spectrum shows double-$\varepsilon$ spacing between the nearest
$\mu$-subband centers, see yellow parts in Fig.~\ref{Fig3}.

\underline{{\bf Near the bottom of els-band}}. For 
${\bar E}-E^{\rm b}_{\rm els}<<1$ and $eV>E^0_{\rm bw}$, 
\begin{equation}\label{16}
E_{n,\mu} =2 -\varepsilon(\mu-5/4)-2\cos k_n, \quad 
\Delta_{\rm c}= \varepsilon,
\end{equation}
where $\mu$ is of the order of $N^{\rm els}$.  Again, we come to  the Wannier quantization rule in two dimensions that refers now not to WSs- but to a part of the upper els-band.

Only small corrections distinguish the results presented above from those which
follows from the analysis  carried out for even ${\cal N}$ \cite{prb63}. It might be of importance however that in the latter case, the $\mu$-subband centered at zero energy does not exist. The predictions made throughout the discussion are not valid for
 ${\cal N}$ of the order of 10 and less. The  high precision numerical solutions of the eigenvalue problem $D_{\cal N}({\bar E})=0$ confirm all analytic and explicit equations presented throughout the discussion.

Since the Nobel Prize was awarded to von Klitzing for the discovery of the integer quantum Hall effect 20 years ago, not a single word has been added to Hall effect in University Physics \cite{UnPhys}. The use of our results for the explanation of the integer and fractional QHEs (in preparation) suggest a promising  new way to go for future research.

\end{document}